\documentclass[twocolumn,superscriptaddress,showpacs,prl]{revtex4}

\usepackage{graphicx}
\usepackage{delarray}
\usepackage{amsmath}
\usepackage{bm}

\newcommand{\braket}[2]{\langle #1 | #2 \rangle}
\newcommand{\ket}[1]{|  #1 \rangle}
\newcommand{\bra}[1]{ \langle #1   |}

\begin{document}
\title{Quantum catalysis of information
}

\author{Koji Azuma}
\email{azuma@qi.mp.es.osaka-u.ac.jp}
\affiliation{Division of Materials Physics, Department of Materials Engineering Science, Graduate School of Engineering Science, Osaka University, 1-3 Machikaneyama, Toyonaka, Osaka 560-8531, Japan}

\author{Masato Koashi}
\affiliation{Division of Materials Physics, Department of Materials Engineering Science, Graduate School of Engineering Science, Osaka University, 1-3 Machikaneyama, Toyonaka, Osaka 560-8531, Japan}
\affiliation{CREST Photonic Quantum Information Project,
4-1-8 Honmachi, Kawaguchi, Saitama 331-0012, Japan}

\author{Nobuyuki Imoto}
\affiliation{Division of Materials Physics, Department of Materials Engineering Science, Graduate School of Engineering Science, Osaka University, 1-3 Machikaneyama, Toyonaka, Osaka 560-8531, Japan}
\affiliation{CREST Photonic Quantum Information Project,
4-1-8 Honmachi, Kawaguchi, Saitama 331-0012, Japan}

\date{\today}

\begin{abstract}
Heisenberg's uncertainty principle and recently derived many no-go theorems including the no-cloning theorem and the no-deleting theorem have corroborated the idea that we can never access quantum information without causing disturbance. Here we disprove this presumption by predicting a novel phenomenon, `quantum catalysis of information,' where a system enables an otherwise impossible task by exchanging information through a quantum communication channel. This fact implies that making use of quantum information
does not always cause disturbance. 
\pacs{03.67.-a, 03.67.Hk, 03.65.Ta}
\end{abstract}

\maketitle

About 80 years ago, Heisenberg implied a groundbreaking notion that measurement on a quantum system inevitably disturbs its state, in contrast to classical systems. This suggests that our accessibility to the information contained in a quantum system is severely limited if we are to keep its state unchanged. In fact, the no-cloning theorem \cite{WZ82,D82} and subsequent no-go theorems \cite{Y86,BBM92,BCFJS96,KI98,KI02} as well as the no-deleting theorem \cite{J02,PB00,Z00,PB00+} have corroborated the idea that we can never access quantum information of a system without causing disturbance.

\begin{figure*}[t]
  \begin{center}
    \includegraphics[keepaspectratio=true,width=12.4cm]{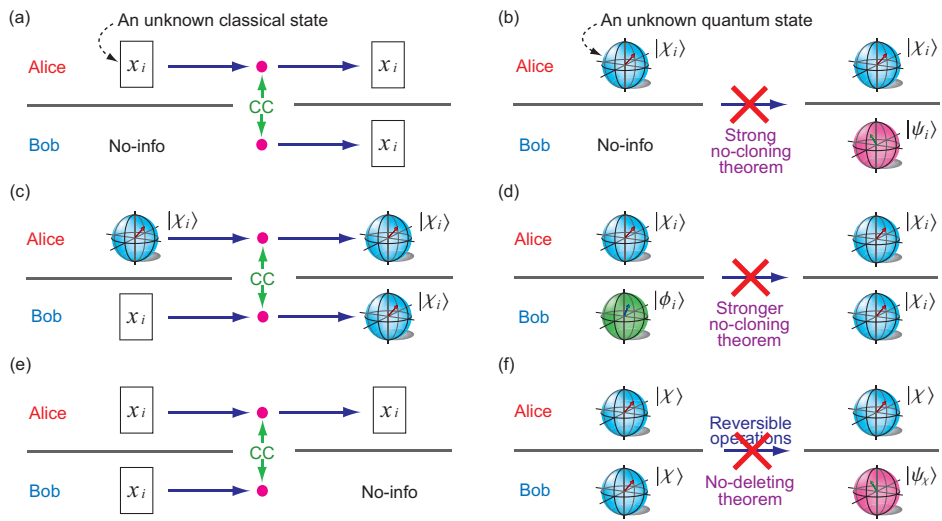}
  \end{center}
  \vspace{-3mm}
  \caption{
Schematic pictures of known processes regarded as catalysis of information, and
 of no-go theorems. 
In each figure, Bob tries to achieve a
 conversion through the help of Alice's catalyst of information, represented by classical state $x_i$ or 
 quantum state $\ket{\chi_i}_A$.
 (a) Cloning of an unknown classical state $x_i$. The task is
 achievable by permitting classical communication (CC) between Alice and
 Bob. 
 (b) Impossibility of extracting information from an
 irreducible set $\{\ket{\chi_i}_A\}$ \cite{BBM92}. 
(c) Cloning with supplementary information \cite{J02}. 
Bob may create a copy of Alice's quantum state $\ket{\chi_i}_A$ 
by using CC with Alice \cite{exBB84}. 
(d) Impossibility of cloning a pair-wise nonorthogonal set
 $\{\ket{\chi_i}_A\}$ with supplementary information
 $\{\ket{\phi_i}_B\}$ \cite{J02}.
 (e) Deletion of an unknown classical state
 $x_i$ \cite{PB00,Z00,PB00+}. Bob can erase his information reversibly through CC with Alice. 
 (f) Impossibility of deletion of a completely unknown quantum state
 $\ket{\chi}_B$ \cite{PB00,Z00,PB00+}. $\ket{\psi_{\chi}}_B$ represents a state from which Bob alone
 cannot resurrect $\ket{\chi}_B$. }
  \label{fig:1}
\end{figure*}

The no-cloning theorem \cite{WZ82,D82,Y86} is well-known as a principle showing a striking difference between the classical and the quantum world.
This theorem states that an unknown quantum state cannot be cloned, in contrast to the classical states that can be freely cloned (Fig.~1a). This prohibition has been strengthened in the so-called strong no-cloning theorem \cite{BBM92}, which asserts that one can extract absolutely no information from an unknown quantum state without disturbing it, let alone create a copy (Fig.~1b). 
If we borrow the concept of `catalysis' in chemical reaction,
the original system in the task of cloning will be regarded 
as a catalyst, which must remain intact 
after it is used as a source of information.
These no-go theorems will thus be understood as implying 
that an unknown quantum state never acts as `a catalyst of information.'

For more general cases where the quantum state is not completely
unknown but was chosen from a set of pure states $\{\ket{\chi_i}\}$,
we can easily find an example of the catalysis.
Suppose that the set $\{\ket{\chi_i}\}$ can be divided into two sets 
$S_1$ and $S_2$ such that 
every state in $S_1$ is orthogonal to any state in $S_2$.
Whenever such reduction is possible, one can apply a projective measurement to learn whether the state $\ket{\chi_i}$ belongs to $S_1$ or $S_2$, without causing any disturbance. Thus, this case is an example in which the state $\ket{\chi_i}$ acts as catalyst of information. However, information extracted from the catalyst is just purely classical. To see this more formally in terms of quantum information theory, we consider a scenario in which the owner of the catalyst, Alice, is communicating with a beneficiary, Bob. 
The above example is then regarded as a `classical' catalysis of information, in the sense that only classical communication is required to transmit the outcome of the projective measurement. 
On the other hand, if the set $\{ \ket{\chi_i}\}$ is irreducible to any two orthogonal sets, the strong no-cloning theorem (Fig.~1b) holds, namely, there is no catalysis. Hence, only classical catalysis is possible as long as Bob initially has no information on Alice's state. This is also true \cite{BCFJS96,KI98,KI02} even when Alice's state is chosen from mixed states.

There are evidences suggesting that the catalysis of information is exclusively classical, even if we allow Bob to have partial information on Alice's state. 
Jozsa considered \cite{J02} the possibility for Alice's catalyst $\{\ket{\chi_i}\}$ to help Bob convert the partial information to an exact clone $\ket{\chi_i}$, and proved \cite{J02} that there is no catalysis if $\{\ket{\chi_i}\}$ includes no orthogonal pair (the `stronger' no-cloning theorem, Fig.~1d). The requisite here is more restrictive than the irreducibility of the set $\{\ket{\chi_i}\}$ mentioned above, and there are cases \cite{J02,exBB84} where an irreducible set $\{\ket{\chi_i}\}$ works as a catalyst (Fig.~1c). We see that classical communication suffices in this case, namely, this is yet another example of classical catalysis of information.

In contrast to the above problems where Bob tries to extract
information from Alice's system, a task with the opposite goal---called deletion of information---has also been considered \cite{PB00,Z00,PB00+,J02}.
In this task, Bob initially has a copy of Alice's state and he wishes 
to erase the information without discarding any subsystem, namely, 
only using reversible operations. Alice's state must be unaltered at the 
end of the process. This scenario can also be regarded as catalysis,
where the role of Alice's catalyst is to absorb the information in Bob's
system. Even for such deletion of information,
the situation looks similar to that of extraction of information:
if the initially shared states are classical,
Bob can completely delete his information \cite{PB00,Z00,PB00+},
which is regarded as classical catalysis (Fig.~1e).
On the other hand, if the initial states are completely unknown quantum 
states, we have the no-deleting theorem \cite{PB00,Z00,PB00+} prohibiting
such catalytic absorption of information (Fig.~1f).

The known results summarized in Fig.~1, including various no-go theorems
 and examples of classical catalysis, corroborate a conventional
 presumption that quantum information in a system cannot 
be accessed without disturbing the system, unlike classical information.
In this Letter, however, we exhibit striking examples of catalysis, which
urges us to abandon such a conventional concept. 
In our examples, Alice's system can help Bob make a clone or delete
 information partially. But unlike the previous examples, 
we can prove that 
her system works as a catalyst only when 
the interaction between Alice and Bob is strong enough to 
allow faithful transmission of an arbitrary state of a qubit,
namely, when they are effectively communicating over 
an ideal quantum channel. Hence the flow of information 
between Alice and Bob is hardly regarded as classical, 
motivating us to call them quantum catalysis of information (Fig.~2).

In order to show that quantum catalysis can be accomplished, we first
assign Bob a task that is not achievable by himself, as in Fig.~2a.
For clarity, we introduce an additional party whom we call Claire. 
She secretly chooses a number $i$ from three candidates
$\{1,2,3\}$.
According to the number $i$, Bob's system $B$ is initially prepared
in state $\ket{\phi_i}_B$, defined by
\begin{equation}
\begin{split}
\ket{\phi_1}&=\ket{0},   \\ 
\ket{\phi_2}&= \ket{0},   \\
\ket{\phi_3}&= \ket{+},\label{eq:sup1'}
\end{split}
\end{equation}
where $\ket{+} \equiv (\ket{0}+\ket{1})/\sqrt{2}$. Bob's goal is to 
convert this state to the corresponding target state 
$\ket{\psi_i}_B$ defined by
\begin{equation}
\begin{split}
   \ket{\psi_1} &=\ket{0},       \\
   \ket{\psi_2} &=\ket{1},       \\
   \ket{\psi_3} &=\ket{+}. \label{eq:ori3'}
\end{split}
\end{equation}
Of course, he can never accomplish this conversion by himself
because states $\ket{\phi_1}_B$ and $\ket{\phi_2}_B$ are identical 
whereas state $\ket{\psi_1}_B$ is different from state $\ket{\psi_2}_B$.

Next, we show that Alice, who already possesses her own target state 
$\ket{\psi_i}_A$, can help Bob to accomplish the process 
$\ket{\phi_i}_B \to \ket{\psi_i}_B$ without disturbing her system.
For example, if Alice and Bob get together and 
apply a controlled-NOT gate \cite{NC} between system $A$ in state
$\ket{\psi_i}_A$ (as control) and system $B$ in state $\ket{\phi_i}_B$
(as target), we obtain state $\ket{\psi_i}_A\ket{\psi_i}_B$ as a result.
This fact assures that conversion $\ket{\psi_i}_A \ket{\phi_i}_B \to
\ket{\psi_i}_A \ket{\psi_i}_B$ is achievable.
Because Alice's system does not receive any disturbance in
this conversion, it is regarded as an example of catalysis of information.

\begin{figure}[b]
  \begin{center}
    \includegraphics[keepaspectratio=true,width=70mm]{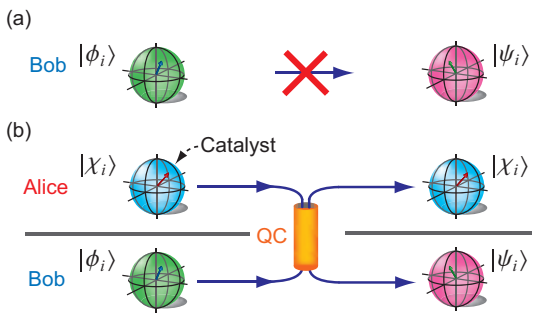}
  \end{center}
  \caption{Schematic pictures of quantum catalysis of information. (a) Bob cannot convert his state $\ket{\phi_i}_B$ into his target state $\ket{\psi_i}_B$ by himself. (b) But he can achieve the conversion by exchanging quantum information with Alice's catalyst $\ket{\chi_i}_A$ through a quantum communication channel (QC).
}
  \label{fig:2}
\end{figure}

Finally, to show that the process $\ket{\psi_i}_A \ket{\phi_i}_B \to
\ket{\psi_i}_A \ket{\psi_i}_B$ is quantum catalysis of information as in  Fig.~2b, 
we prove that, if Alice and Bob can achieve the catalysis, they have the potential to freely exchange an unknown state of a qubit between them.
Note that, at first glance, there is no apparent evidences to prove this, because both of the input $\ket{\psi_i}_A \ket{\phi_i}_B$ and the output $\ket{\psi_i}_A \ket{\psi_i}_B$ of the catalysis $\ket{\psi_i}_A \ket{\phi_i}_B \to \ket{\psi_i}_A \ket{\psi_i}_B$ are separable states.
In the proof below, we consider a choice of another fictitious
input state and 
ask what the output state would be, by
invoking a property of the disturbance-free
operations \cite{KI98,KI02} which has been studied in quantum information theory.

According to quantum mechanics, any physical process applied to a system $S$ can be described \cite{NC} by a unitary operation $\hat{U}$
on the system $S$ and an auxiliary system $E$ prepared in a fixed state $\ket{\Sigma}_E$.
Hence, the process $\ket{\psi_i}_A \ket{\phi_i}_B \to \ket{\psi_i}_A \ket{\psi_i}_B$ should be described by
\begin{eqnarray}
\hat{U}\ket{\psi_i}_A \ket{\phi_i}_B \ket{\Sigma}_E= \ket{\psi_i}_A \ket{\psi_i}_B \ket{\Sigma_i}_E, \label{eq:u}
\end{eqnarray}
where $\ket{\Sigma_i}_E$ is the final state of auxiliary system $E$, which might depend on the identity of the input state $\ket{\psi_i}_A \ket{\phi_i}_B$ in general.
However, for states $\{\ket{\phi_i}\}$ and $\{ \ket{\psi_i} \}$ of Eqs.~(\ref{eq:sup1'}) and (\ref{eq:ori3'}), the state $\ket{\Sigma_i}_E$ should be independent of the identity of the input $\ket{\psi_i}_A \ket{\phi_i}_B$, namely, $\ket{\Sigma_1}_E=\ket{\Sigma_2}_E=\ket{\Sigma_3}_E$;
this fact can be easily shown by taking the inner product between Eq.~(\ref{eq:u}) with $i=j$ $(j=1,2)$ and Eq.~(\ref{eq:u}) with $i=3$, and by noting
$
\braket{\phi_j}{\phi_3}=\braket{\psi_j}{\psi_3}=1/\sqrt{2}\; (j=1,2). 
$
Such independence of the auxiliary system $E$ assures that the process
$\ket{\psi_i}_A \ket{\phi_i}_B \to \ket{\psi_i}_A \ket{\psi_i}_B$ is a
coherence-preserving operation for any state in the Hilbert subspace spanned by three states $\{\ket{\psi_i}_A \ket{\phi_i}_B\}$: if Alice and Bob execute the process $\ket{\psi_i}_A \ket{\phi_i}_B \to \ket{\psi_i}_A \ket{\psi_i}_B$ with an initial state in the form of $\sum_{i=1}^3 \alpha_i \ket{\psi_i}_A \ket{\phi_i}_B$, they should obtain coherent superposition $\sum_{i=1}^3 \alpha_i \ket{\psi_i}_A \ket{\psi_i}_B$ as the output.

Let us consider a case where Alice and Bob start from an input state described by 
$
\ket{\Psi^{\rm in}}_{AB} = \frac{1}{\sqrt{2}} \sum_{i=1}^2 \ket{\psi_i}_A \ket{\phi_i}_B= \ket{+}_A\ket{0}_B.
$
Then, from the fact in the previous paragraph, the output state of the catalysis $\ket{\psi_i}_A \ket{\phi_i}_B \to \ket{\psi_i}_A \ket{\psi_i}_B$ should be 
$
\ket{\Psi^{\rm out}}_{AB} = \frac{1}{\sqrt{2}} \sum_{i=1}^2 \ket{\psi_i}_A \ket{\psi_i}_B  = \frac{1}{\sqrt{2}} (\ket{0}_A \ket{0}_B + \ket{1}_A \ket{1}_B).
$
Note that the input state $\ket{\Psi^{\rm in}}_{AB}$ is a separable state, but the output state $\ket{\Psi^{\rm out}}_{AB}$ is an Einstein-Podolsky-Rosen (EPR) pair \cite{EPR,B51}.
This means that, if Alice and Bob can achieve the process
$\ket{\psi_i}_A \ket{\phi_i}_B \to \ket{\psi_i}_A \ket{\psi_i}_B$, the interaction between them must have the potential to generate an EPR pair from a separable state.
With classical communication, Alice and Bob could use the 
EPR pair to transmit an unknown state of a qubit faithfully
between them, using the protocol of quantum teleportation \cite{BBCJPW93}.
Thus, the interaction between Alice and Bob required to achieve the
catalysis $\ket{\psi_i}_A \ket{\phi_i}_B \to \ket{\psi_i}_A
\ket{\psi_i}_B$ is strong enough to allow faithful transmission of an
arbitrary state of a qubit, which urges us to interpret the flow of
information in the catalysis to be genuinely quantum one.
Therefore, the process $\ket{\psi_i}_A \ket{\phi_i}_B \to \ket{\psi_i}_A \ket{\psi_i}$ is an example of quantum catalysis of information.

The above quantum catalysis 
{can be regarded as a kind of cloning task}
because Bob's goal is making the copy of Alice's information.
Interestingly, we can also find quantum catalysis in a task of deletion.
Let us consider the inverse to the above catalysis, namely, $\ket{\psi_i}_A \ket{\psi_i}_B \to \ket{\psi_i}_A \ket{\phi_i}_B$.
This process may be regarded as a kind of deletion in the following viewpoints:
(i) the process is reversible, because it can be accomplished by a controlled-NOT gate between system $A$ in initial state $\ket{\psi_i}_A$ (as control) and system $B$ in initial state $\ket{\psi_i}_B$ (as target);
(ii) after the process $\ket{\psi_i}_A \ket{\psi_i}_B \to \ket{\psi_i}_A \ket{\phi_i}_B$, Bob can never resurrect state $\ket{\psi_i}_B$ from $\ket{\phi_i}_B$ by himself.
In what follows, we show that this deletion process is also quantum catalysis of information.
In order to prove this, we consider a case where Alice and Bob conduct the catalysis $\ket{\psi_i}_A \ket{\psi_i}_B \to \ket{\psi_i}_A \ket{\phi_i}_B$ for the following input:
\begin{align}
\ket{\Phi^{\rm in}}_{AB} =& \frac{1-i}{2}  \ket{\psi_1}_A \ket{\psi_1}_B -  \frac{1+i }{2}  \ket{\psi_2}_A \ket{\psi_2}_B \nonumber \\ &+ i \ket{\psi_3}_A \ket{\psi_3}_B \nonumber \\
=& \frac{1}{2} (\ket{0}_A+i \ket{1}_A)(\ket{0}_B+i \ket{1}_B), \label{Phi_in}
\end{align}
where the input $\ket{\Phi^{\rm in}}_{AB}$ is chosen to be in the Hilbert space spanned by states $\{\ket{\psi_i}_A \ket{\psi_i}_B\}$.
Similarly to the proof for the previous quantum catalysis $\ket{\psi_i}_A \ket{\phi_i}_B \to \ket{\psi_i}_A \ket{\psi_i}_B$, we can show that the output should be coherent superposition in the form of
\begin{align}
\ket{\Phi^{\rm out}}_{AB} =& \frac{1-i}{2}  \ket{\psi_1}_A \ket{\phi_1}_B -  \frac{1+i}{2}  \ket{\psi_2}_A \ket{\phi_2}_B \nonumber \\
&+ i \ket{\psi_3}_A \ket{\phi_3}_B \label{Phi_out}  \\
=& \frac{1}{\sqrt{2}} (\ket{-}_A \ket{0}_B + i \ket{+}_A \ket{1}_B),
\end{align}
where $\ket{-} \equiv (\ket{0}-\ket{1})/\sqrt{2}$.
Because $\ket{\Phi^{\rm in}}_{AB}$ is a separable state but
$\ket{\Phi^{\rm out}}_{AB}$ is an EPR pair, if Alice and Bob can achieve
the catalysis $\ket{\psi_i}_A \ket{\psi_i}_B \to \ket{\psi_i}_A
\ket{\phi_i}_B$, they must have the ability to accomplish faithful
transmission of an unknown state of a qubit between them.
Hence, the process $\ket{\psi_i}_A \ket{\psi_i}_B \to \ket{\psi_i}_A \ket{\phi_i}_B$ is quantum catalysis of information.
Note that there is no reason why the state left after the deletion must be $\ket{\phi_i}_B$, in order to satisfy the 
conditions (i) and (ii).
Hence one may ask whether a deleting process
leaving a different state $\ket{\phi'_i}_B$ can be regarded as a
 classical one.  
Interestingly, with any choice of $\ket{\phi'_i}_B$
that satisfies (i) and (ii), the process
$\ket{\psi_i}_A \ket{\psi_i}_B \to \ket{\psi_i}_A \ket{\phi_i'}_B$
can be shown to be quantum catalysis of information.
First note that $\ket{\phi'_i}_B$ must satisfy
$\braket{\phi_1'}{\phi_3'}=\braket{\phi_2'}{\phi_3'}=1/\sqrt{2}$ 
from the reversibility of the deletion.
If we assume that the separable state, $\ket{\Phi^{\rm in}}_{AB}$
of Eq.~(\ref{Phi_in}), is chosen as the input of the deletion process,
the output state $\ket{\Phi'^{\rm out}}_{AB}$ can be evaluated in a similar manner by replacing $\ket{\phi_i}_B$ with $\ket{\phi'_i}_B$ in Eq.~(\ref{Phi_out}).
One can easily confirm that ${}_B\bra{\phi_1'} \ket{\Phi'^{\rm out}}_{AB}$ and ${}_B\bra{\phi_3'} \ket{\Phi'^{\rm out}}_{AB}$ refer to the same state (up to normalization) of system $A$ only if $\braket{\phi_1'}{\phi_2'} = 0$.
The output $\ket{\Phi'^{\rm out}}_{AB}$ is thus entangled 
whenever (ii) is satisfied, namely whenever $\braket{\phi_1'}{\phi_2'}
\neq 0$.
Hence, every deletion process $\ket{\psi_i}_A \ket{\psi_i}_B \to
\ket{\psi_i}_A \ket{\phi_i'}_B$ should be regarded as quantum catalysis of information.

The discovery of the quantum catalysis of information implies the inequivalence of the
two plausible characterizations of the quantum information as opposed to 
the classical information.
One is the clear definition provided during the recent development of  
the quantum information theory: quantum information is one that cannot be conveyed without quantum communication.
Our choice of the term `quantum catalysis' comes from 
adoption of this modern definition through the discussion in this paper.
The other characterization dates back to the no-cloning
theorem \cite{WZ82,D82,Y86}, or even to Heisenberg: information that cannot be
accessed without disturbing the state of the system. Let us call
this type of information `fragile information' for the moment,
to clarify the distinction from the former one.
As seen in  Fig.~1, 
subsequently found no-go theorems such as the strong no-cloning
theorem \cite{BBM92}, the no-deleting theorem \cite{PB00,Z00,PB00+,J02} and
the stronger no-cloning theorem \cite{J02} supported our expectation 
that the quantum information would be indeed the fragile information, 
and catalysis would only be allowed for the classical information that 
can be exchanged through classical communication. 
However, quantum catalysis of information shown here implies that 
the two characterizations do not coincide, namely,
without receiving disturbance, a system may exchange a type of
information that can only be transmitted through 
quantum communication channels.
Furthermore, the fact that the quantum catalysis is found not only in a
situation like cloning but also in a situation corresponding to deletion
suggests that there may be many such examples, and 
the boundary between the classical and the quantum world 
is far more complicated than the existing no-go theorems suggest.

Finally, we raise several open questions deserving further intensive studies. 
Existence of quantum catalysis of information shown here is only the first step for grasping complicated nature of quantum information.
In fact, we hardly understand what properties of quantum states permits 
quantum catalysis. 
Solving this problem is essential for precise characterization of  
what the fragile information really is.
A quantitative understanding of quantum catalysis is also worth
investigating.
In the examples we found, one EPR pair is shown to be necessary
and sufficient for achieving the quantum catalysis of information by
classical communication. Since sending Alice's state to Bob and then
back to Alice requires two EPR pairs, there might be a 
qubit quantum catalyst requiring two EPR pairs, or there might be 
a general bound on this cost.
A more complicated scenario in which the catalyst itself is composed 
of a bipartite system with LOCC constraint also has its own interest 
in conjunction with the entanglement theory. Whereas existence 
of catalysis in such a scenario has been found \cite{JP99ec}, nature of 
the interaction between the catalyst and the beneficiary still awaits 
further investigation.
We believe that clearer understanding of quantum catalysis of
information will help us grasp the true nature of quantum information,
and may lead to novel applications of information processing.

We thank N. Nagaosa, S. Murakami, T. Yamamoto,  J. Shimamura, R. Namiki, S. K. Ozdemir, T. Ohnishi, and S. Kuga for valuable discussions.
This work was supported by a MEXT Grant-in-Aid for Young Scientists (B) No.~17740265, and by JSPS Research Fellowships for Young Scientists No.~19$\cdot$1377.


\begin{thebibliography}{99}
\bibitem{exBB84}
For example, suppose that 
Alice's states $\{\ket{\chi_i}_A\}$ are the four states in the BB84 
quantum key distribution protocol \cite{BB84}, and Bob's classical state
$\{x_i\}$ has the basis information: $\ket{\chi_1}_A=\ket{0}_A$,
 $x_1=0$, $\ket{\chi_2}_A=\ket{1}_A$, $x_2=0$,
 $\ket{\chi_3}_A=(\ket{0}_A+\ket{1}_A)/\sqrt{2}$, $x_3=1$,
  $\ket{\chi_4}_A=(\ket{0}_A-\ket{1}_A)/\sqrt{2}$, and $x_4=1$.
\end{thebibliography}
\end{document}